\documentclass[12pt,titlepage]{article}
\usepackage{times}
\usepackage{setspace}
\usepackage{acronym}
\usepackage{amsmath}
\usepackage{amssymb}
\usepackage{graphicx}
\usepackage{hyperref}
\usepackage{multirow}
\usepackage{epstopdf}
\usepackage{subfigure}
\usepackage{rotating}
\usepackage{ulem}
\usepackage[sort&compress]{natbib} 
\bibpunct{[}{]}{,}{s}{}{;}
\DeclareGraphicsExtensions{.ps,.pdf,.eps,.jpg}

\newcommand{\be}{\begin{equation}}
\newcommand{\ee}{\end{equation}}

\newcommand{\magsim}
{\ \lower2pt\hbox{$\sim $}\mkern-14mu \raise2pt\hbox{$>$}\ }

\newcommand{\etal}{\mbox{\rm et al.}}     
\newcommand{\mearth}{M$_{\rm Earth}~$}     

\usepackage{graphicx}

\begin{document}


\begin{titlepage}

\begin{center}

\textbf{{\Large \bf Dynamical Models of Terrestrial Planet Formation }}\\

\vspace{.25in}

\begin{description} 

\item Jonathan I. Lunine$^{\ast}$, Lunar and Planetary Laboratory, The University of Arizona, Tucson AZ USA, 85721. jlunine@lpl.arizona.edu 
\item  David P. O'Brien, Planetary Science Institute, Tucson AZ USA 85719
\item Sean N. Raymond, Center for Astrophysics and Space Astronomy, UCB 389, University of Colorado, Boulder, CO 80309-0389 and Laboratoire d'Astrophysique de Bordeaux (CNRS; UniversitŽ Bordeaux I), BP 89, F-33270 Floirac, France
\item Alessandro Morbidelli, Obs. de la C\^{o}te d'Azur, Nice, F-06304 France 
\item Thomas Quinn, Department of Astronomy, University of Washington, Seattle USA 98195
\item Amara L. Graps, Southwest Research Institute, Boulder CO USA 80302

\end{description}

\vspace{0.1in}

Accepted for publication in Advanced Science Letters June 10, 2009

\vspace{0.1in}

\end{center}

$^{\ast}$Corresponding author

\end{titlepage}
\clearpage

\abstract{ 

We review the  problem of the formation of terrestrial
planets, with particular emphasis on the interaction of dynamical and
geochemical models.  The lifetime of gas around stars in the process of
formation is limited to a few million years based on
astronomical observations, while isotopic dating of meteorites and the
Earth-Moon system suggest that perhaps 50-100 million years were required for the
assembly of the Earth. Therefore, much of the growth of the terrestrial
planets in our own system is presumed to have taken place under largely gas-free conditions, and
the physics of terrestrial planet formation is dominated by gravitational
interactions and collisions.  The earliest phase of terrestrial-planet 
formation involve the growth of km-sized or larger planetesimals from
dust grains, followed by the accumulations of these planetesimals into
 $\sim$100 lunar- to Mars-mass bodies that
are initially gravitationally isolated from one-another in a swarm of smaller
planetesimals, but eventually grow to the point of significantly perturbing
one-another.  The mutual perturbations between the embryos, combined with
gravitational stirring by Jupiter, lead to orbital crossings and collisions
that drive the growth to Earth-sized planets on a timescale of $10^7-10^8$
years.  Numerical treatment of this process
has focussed on the use of symplectic integrators which can rapidy integrate
the thousands of gravitationally-interacting bodies necessary to accurately
model planetary growth.  While the general nature of the terrestrial
planets--their sizes and orbital parameters--seem to be broadly reproduced by
the models, there are still some outstanding dynamical issues. One of these is the presence of an embryo-sized
body, Mars, in our system in place of the more massive objects that
simulations tend to yield. Another is the effect such impacts have on the geochemistry of the growing planets;  re-equilibration of isotopic ratios of major elements during giant impacts (for example) must be considered in comparing the predicted compositions of the terrestrial planets with the geochemical data.  As the dynamical models become successful in reproducing the
essential aspects of our own terrestrial planet system, their utility in
predicting the distribution of terrestrial planet systems around other stars,
and interpreting observations of such systems, will increase.}

\vspace {0.2in}


\vspace{1.0in}

Dedicated to George Wetherill (1925-2006), pioneer in studies of the formation
of the terrestrial planets. 

\newpage

\section[Introduction]{Introduction}

\begin{figure}
\includegraphics[width=5in]{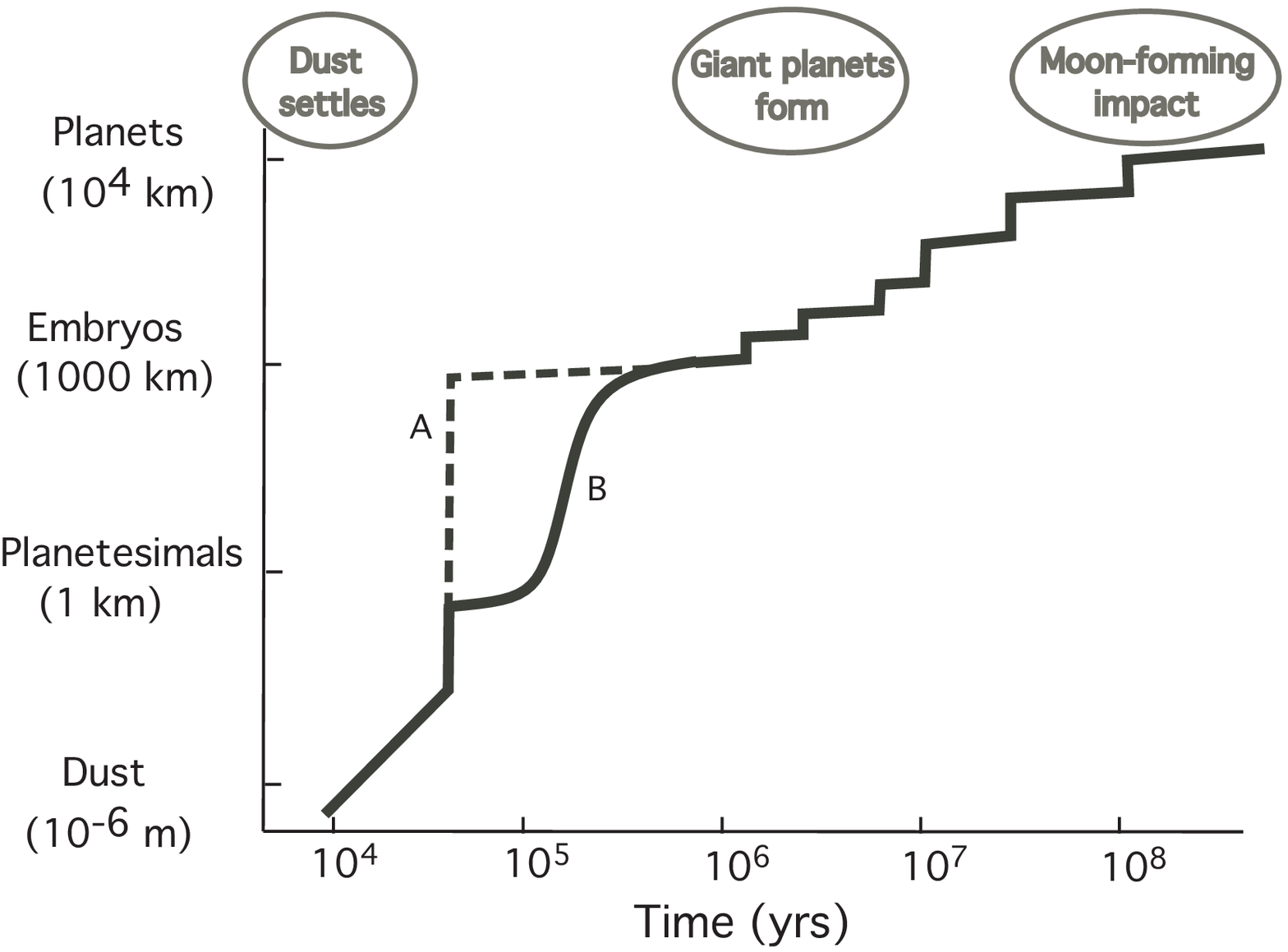} 
\caption{An illustration of the stages of terrestrial planet growth and the
relevant timescales \citep[not to scale; image from][]{Rayb2009}.  See \S 3
for details.}
\label{fig:schema}
\end{figure}

The formation of the terrestrial planets remains one of the enduring problems
in planetary science and (in view of the expectation of large numbers of extrasolar terrestrial-type planets) astrophysics today. The complexity of terrestrial geochemistry,
constraints on timescales, the presence of abundant water on the Earth, and
the curious geochemical and dynamical relationships between the Earth and the
Moon are among the problems that must be addressed by models. 
Pioneering studies by Safronov \citep{Safronov1969} and successors such as  Weidenschilling \citep{Weid76} established the basic physics of gas-free accretion. The effects of gas on accretion were examined somewhat later, most notably by the ``Kyoto" school of Hayashi and collaborators \citep{Haya81}.
In the 1980's, studies of terrestrial planet formation advanced further thanks to George Wetherill
\citep{Wetherill1980}, his students and postdoctoral collaborators, who highlighted the basic problems of obtaining the correct
low planetary eccentricities and inclinations, as well as producing a
diversity of sizes ranging from Earth through Mars and Mercury. Breakthroughs
in the subject came through the development of special numerical approaches to
the problem, as well as theoretical insights that allowed for the right
starting boundary conditions.  Additional geochemical considerations,
including formation timescales derived from radioactive isotopic ratios, and
stable isotopic constraints on source regions, continue to challenge the
models today.
Decades of research have established a rough timeline of events during
the formation of the Solar System's terrestrial planets. These are summarized
in Figure~\ref{fig:schema}, which shows the many steps which occurred during
the formation of the Earth. 

The classical view, developed in the 1960's and 1970's, is that the planetesimals
grow gradually, from collisional coagulation of pebbles and boulders.  The growth becomes exponential (runaway) when the first massive bodies appear
in the disk \citep[]{Greenberg1978,Wetherill1989}. 
However, it is not clear how ordered growth can procede beyond 1 meter in size, the so-called
meter-size barrier that we explain more extensively in section 2. A new view
to by-pass the meter-size barrier is that boulders, pebbels and even
chondrule-size particles can be concentrated in localized structures of a
turbulent disk, where they form self-gravitating clumps. The
size-equivalent of these clumps can be 10km \citep{Goldreich1973}, 100km
\citep{Cuzzi2008} or 1,000 km \citep{Johansen2007}, depending on the models 
and the physics that is accounted for. The growth rate of $\sim$
Moon-sized embryos decreases during oligarchic growth because of viscous
stirring of planetesimals by the embryos and decreased gravitational focusing
\citep{Ida1992a,Ida1992b}.  Late-stage accretion begins when embryo-embryo
collisions occur \citep{Wetherill1985, Kenyon2006}, and takes place in the
presence of Jupiter and Saturn, which must have formed in less than $\sim$ 5
Myr \citep{Haisch2001}.  Late-stage accretion lasted for about 100 Myr in the
Solar System based on radioisotopic chronometers. \citep{Touboul2007}. 

In this review we describe the numerical tools and theoretical concepts used
in simulating terrestrial planet formation, and the geochemical constraints.
We focus on two applications: (1) the origin of water on the Earth and (2) the
predicted diversity of terrestrial planet systems around other stars. We begin
by describing the astrophysical and geochemical constraints on timescales. We
then describe the phases of planetesimal growth and the subsequent oligarchic
growth of planetary embryos that set the boundary conditions for terrestrial
planet formation, following which the numerical approach widely used today is
outlined. We discuss results from the various groups that have conducted
simulations, and how well certain constraints from observations are reproduced. The relevance
of the formation of the Moon by giant impact in understanding terrestrial
planet formation is considered. We then highlight application of the
simulations to the origin of water on the Earth, and to simulation of
extrasolar planetary systems.  We close with a list of outstanding issues, and
the possible directions for their solution.

\section{Early Phases of Terrestrial Planet Formation}

\subsection{Planetesimal Formation}

Planets form from disks created when clumps of interstellar
gas and dust, organized in dense molecular clouds, collapse to form stars
\citep{PPV}. The angular momentum content of typical clumps ensures that a
portion of the collapsing material ends up in a disk, through which much of
the mass works its way inward to the growing ``protostar" and angular momentum
continues to reside in the disk--fully consistent with the mass and angular
momentum distribution of the Sun and planets. Disks undergo evolution from
gas-dominated systems to ``debris" disks in which only solids remain; based on
astronomical observations most of the gas is gone within 6 million years after
the collapse begins \citep{Najita2007}. (The appearance of the first solids in our solar system is reliably dated by meteorites to be 4.568 billion years ago \citep{Moy07}).

``Planetesimal" is the term used to connote the fundamental building blocks of
the planets whose growth is dominated by gravity rather than gas-drag.  They
are generally defined to be the smallest rocky bodies that are decoupled from
the gaseous disk.  The most commonly-assumed planetesimal size is 1 km,
corresponding to a mass on the order of 10$^{16}$ grams.  However, km-sized
bodies are not completely decoupled from the gas, in that their orbits are
significantly altered by gas drag via relatively rapid ($\sim 10^3 - 10^4$ yr)
damping of their eccentricities and inclinations, and much slower ($\sim 10^6$
yr) decay of their semimajor axes \citep{Adachi1976}.  In fact, the actual
size distribution of bodies during the phase of gravity-dominated growth is
determined by the formation mechanism of these bodies, which remains uncertain
(see below).  The planetesimal size is therefore used as a parameter in some
models of later stages of planetary growth \citep{Chambers2006}.

Modeling planetesimal growth requires a detailed treatment of the structure of
the gaseous disk, including turbulence, local pressure gradients, magnetic
processes, and vortices.  Models can be constrained by observations of dust
populations in disks around young stars, although interpretation of
observations remains difficult \citep{Dullemond2005}.  There currently exist
two qualitatively-different theories for planetesimal formation: collisional
growth from smaller bodies \citep[eg.,][]{Weidenschilling1993} and local
gravitational instability of smaller bodies \citep[]{Goldreich1973,
Youdin2002, Johansen2007,Cuzzi2008}.

Collisional growth of micron-sized grains, especially if they are arranged
into fluffy aggregates, appears efficient for relatively small particle sizes
and impact speeds of $\sim 1 \, m\, s^{-1}$ or slower\citep{Wurm2000, Poppe2000, Benz2000}; see the review by Dominik \etal (2007)\citep{Dominik1997}.  However,
there is a constant battle between disk turbulence, which increases random
velocities, and drag-induced settling, which reduces them \citep{Cuzzi1993,
Cuzzi2006}.  Growth of particles in such collisions appears effective until
they reach roughly 1 cm to 1 m in size.  At that point, continued growth may
be suppressed by collision velocities of $\geq 10 \, m\, s^{-1}$
\citep{Dullemond2004, Dullemond2005}.

Meter-sized bodies are the barrier of planetesimal formation.  As an object
in the gaseous disk grows, it becomes less strongly-coupled to the gas such
that its orbital velocity transitions between the gas velocity, which is
slowed by partial pressure support, and the local Keplerian velocity.  This
increases the relative velocity between the object and the local gas such that
the object feels a head wind which acts to decrease its orbital energy and
cause the body to spiral toward the star.  Large ($\geq$ 10s to 100s of
meters) objects have enough inertia that orbital decay occurs slowly, but
there exists a critical size for which orbital decay is fastest.  For the case
of rocky bodies in a gaseous disk, this critical size is roughly 1 m, and the
timescale for infall for meter-sized bodies can be as short as 100 years.  This is referred to as the `meter-size
``catastrophe" or sometimes ``barrier", because the infall timescale is far shorter than typical growth
timescales \citep{Weidenschilling1977b}. Collisional growth models must therefore quickly cross the barrier
at meter-sizes if they are to reach planetesimal sizes
\citep{Weidenschilling1993, Benz2000, Weidenschilling2000}.

The gravitational instability model for planetesimal formation suggests that a
large number of small patches of particles could become locally
gravitationally unstable and form planetesimals \citep{Safronov1969, Goldreich1973,
Youdin2002}.  (The criterion for gravitational instability of Keplerian disks appears already in Safranov (1960)) \citep{Safronov1960}. This process requires a concentration of meter-sized or smaller
particles.  If the density of solids in a small patch exceeds a critical
value, then local gravitational instability can occur, leading to top-down
formation of planetesimals.  A concentration of small particles great by a
large factor compared with the gas is the key to the process.

Models for the concentration of small particles often rely on structure within
the gaseous component of the disk, generated by turbulence or self-gravity
\citep{Cuzzi2001,Rice2006}.  If the disk is even weakly turbulent, a
size-dependent concentration of small particles can occur \citep{Cuzzi1996,Cuzzi2001,Cuzzi2006}. Pre-existant
chondrule-sized particles may have been concentrated at these scales by such a mechanism, thus appearing as the basic building blocks of larger structures such as
chondritic parent bodies. 

\begin{figure}[t]
\begin{center}
\includegraphics[width=5in]{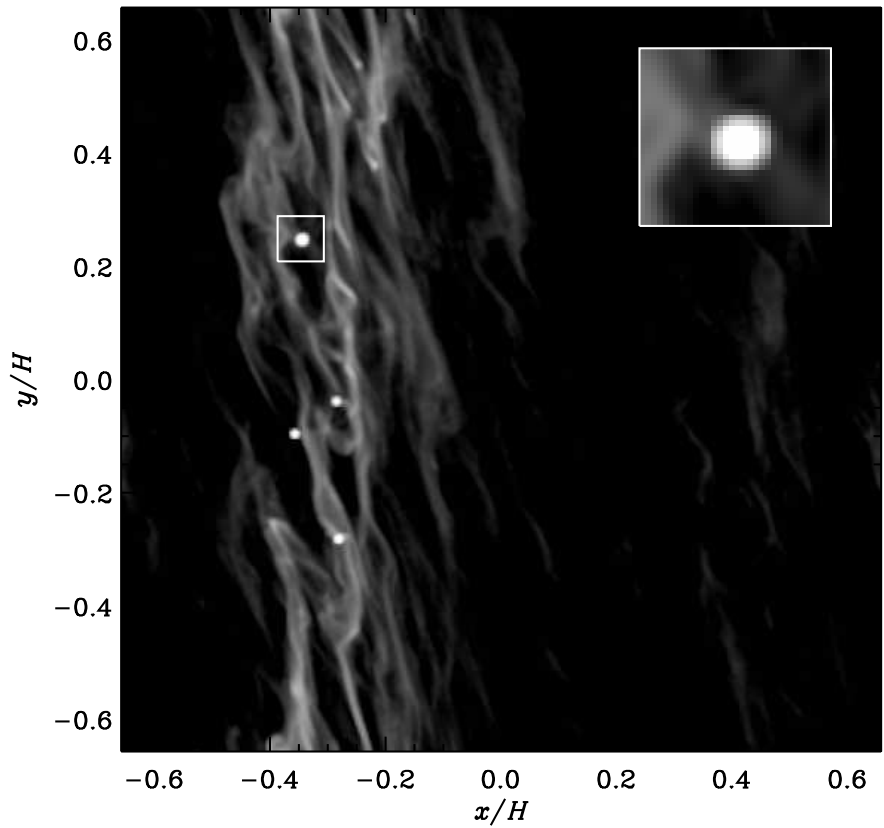} 
\caption{Concentration of boulder-sized particles in MRI-turbulent structures
in the model of Johansen et al. (2007) \citet{Johansen2007}.  The x and y axes are shown in units
of the disk's vertical scale height H, and this snapshot is from seven orbital
times after a clumping event occurred.  The greyscale represents the local
density of particles, and the solid circles show the location of four clumps
that are each more massive than Ceres (i.e., they correspond to $\sim$ 1000 km
or larger ``planetesimals" (or small embryos) in the overdense filament.  The
inset focuses on one clump as shown. Original figure provided by Anders Johansen.}
\label{fig:pllform}
\end{center}
\end{figure}

Self-gravitating clumps of chondrules may end up as 10- to 100-km sized
planetesimals; in this case particles don't collapse rapidly on the dynamical
timescale but slowly contract into planetesimals \citep{Cuzzi2008}. Turbulence
can also concentrate larger, meter-sized particles by producing local pressure
maxima which can act as gathering points for small bodies.  As for the
meter-size catastrophe, boulder-sized objects are the fastest to drift toward
pressure maxima \citep{Haghighipour2003, Johansen2006}.\footnote{In fact, the
idea of the meter-sized catastrophe assumes that the disk has a smooth
pressure gradient \citep{Weidenschilling1977b}.  For disks with small-scale
pressure fluctuations, small particles do not necessarily spiral inward but
simply follow the local pressure gradient \citep{Haghighipour2003}.}  The
concentration in these regions can be further increased via a streaming
instability between the gas and solids \citep{Youdin2005, Johansen2007b}, and
gravitational collapse of the clumps can occur in these dense regions.
Johansen et al.\citet{Johansen2007} showed that planetesimals can form via this process and
that the particle clumps (i.e., the rubble-pile planetesimals) have a
distribution of sizes that ranges up to 1000 km or larger.
Figure~\ref{fig:pllform} shows the surface density of boulder-sized particles
in a disk in which
four 1000 km-scale objects have formed \citep{Johansen2007}.  An alternate location for
planetesimal formation via gravitational instability are regions with an
increased local density of solids \citep{Goodman2000}.  Other ways to
concentrate solids include drag-induced in-spiralling to disk edges (\citep{Youdin2002}, 
vortices \citep{Tanga1996, Barge1995}, or photo-evaporative depletion of
the gas layer \citep{Throop2005}.

\subsection[Oligarchic growth]{Oligarchic growth}

Relative velocities in the disk temd to remain low, whether because of damping of eccentricities by gas drag \citep{Adachi1976}, collisional damping, or merely the presence of a few larger bodies that can limit the dispersion velocities of the smaller ones. Bodies that are slightly larger than the typical size can increase their
collisional cross sections due to gravitational focusing and thereby
accelerate their growth \citep{Safronov1969,Greenberg1978}:


\begin{equation} \frac{dM}{dt} = \frac{\pi R^2 \Sigma \, v_{rand}}{2 \, H}
 F_g(v_{rand}),
\end{equation} 

where $R$ represents the body's physical radius,
$v_{rand}$ represents the velocity dispersion of planetesimals, $\Sigma$ is
the local surface density of planetesimals, $H$ is the scale height of the
planetesimal disk, and $F_g$ is the so-called gravitational focussing factor,  which depends on $V_{rand}$.  The expression for $F_g$ is complicated \citep{Greenzweig1992, Greenzweig1990}; for moderate $v_{rand}$, it can be approximated by
$\left(1+\frac{v_{esc}2}{v_{rand}2}\right)$, where $v_{esc}$
is the escape speed from the body's surface.

While random velocities are small, gravitational focusing
can increase the growth rates of bodies by a factor of hundreds, such that
$dM/dt \sim M^{4/3}$, leading to a phase of rapid ``runaway growth"
\citep{Safronov1972, Greenberg1978, Wetherill1989, Wetherill1993, Ida1992a, Ida1992b,
Kokubo1996, Goldreich2004}. The length of this phase depends on the timescale
for $v_{rand}$ to increase, which depends on a combination of eccentricity
growth via interactions with large bodies and eccentricity damping. For small
($\sim$100 m-sized) planetesimals, gas drag is stronger such that runaway
growth can be prolonged and embryos may be larger and grow faster
\citep{Rafikov2003, Chambers2006}.

As large bodies undergo runaway growth, they gravitationally perturb
nearby planetesimals. The random velocities of planetesimals are therefore
increased by the larger bodies in a process called ``viscous stirring"
\citep{Ida1992a}. During this time, the random velocities of large bodies are
kept small via dynamical friction with the swarm of small bodies
\citep{Ida1992b}. As random velocities of planetesimals increase,
gravitational focusing is reduced, and the growth of large bodies is slowed to
the geometrical accretion limit, such that $dM/dt \sim M^{2/3}$
(\citep{Ida1993, Rafikov2003}).  Nonetheless, large bodies continue to grow, and
jostle each other such that a characteristic spacing of several mutual Hill
radii $R_{H, m}$ is maintained ($R_{H,m} \equiv 0.5
[a_1+a_2]\,[M_1+M_2/3M_\star]^{1/3}$, where $a_1$ and $M_1$ denote the orbital
distance and mass of object 1, etc. \citep{Kokubo1995}).  This phase of
growth is often referred to as ``oligarchic growth", as just a few large bodies
dominate the dynamics of the system, with reduced growth rates and increased
interactions between neighboring embryos \citep{Kokubo1998, Kokubo2000,
Kokubo2002, Leinhardt2005}.

\begin{figure}
\includegraphics[width=6in]{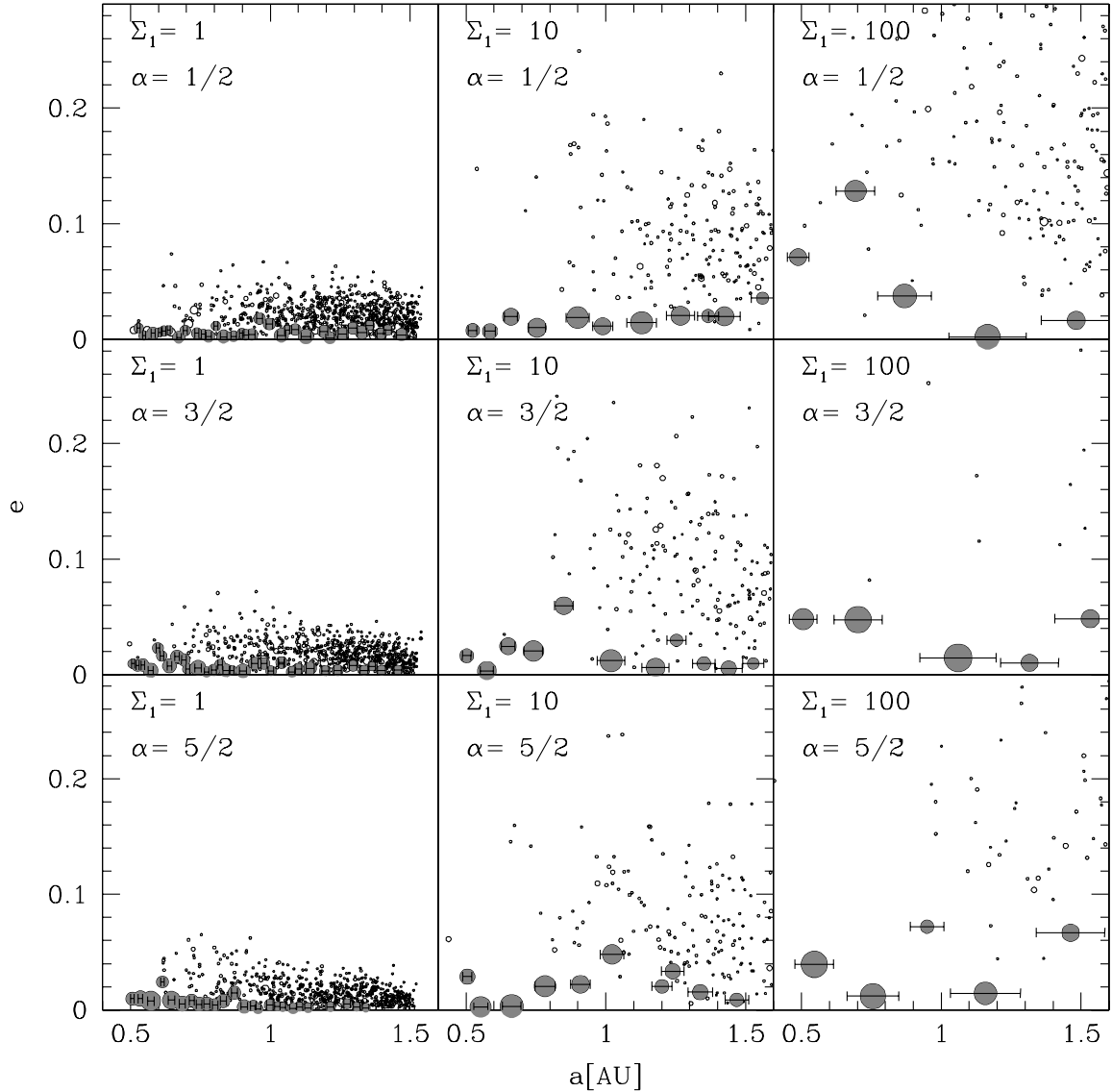}
\caption{Figure 14 from Leinhardt \& Richardson (2005)\citep{Leinhardt2005} showing their
distributions of embryos and smaller bodies with a range in surface density exponents $\alpha$
and surface densities $\Sigma_1$ . All panels are at 500,000 years, except for panel 3 in rows 1 and 3 which are at 110,000 and  225,000 years, respectively. The horizontal bars represent 10 times the Hill radii.This figure was cited from Leinhardt \& Richardson,  Astrophys. J., 625, 427 (2005) with copyright permission
from IoP publishers.}.  
\label{fig:leinhardt_fig2}
\end{figure}

\begin{figure}[t]
\begin{center}
\includegraphics[width=3in]{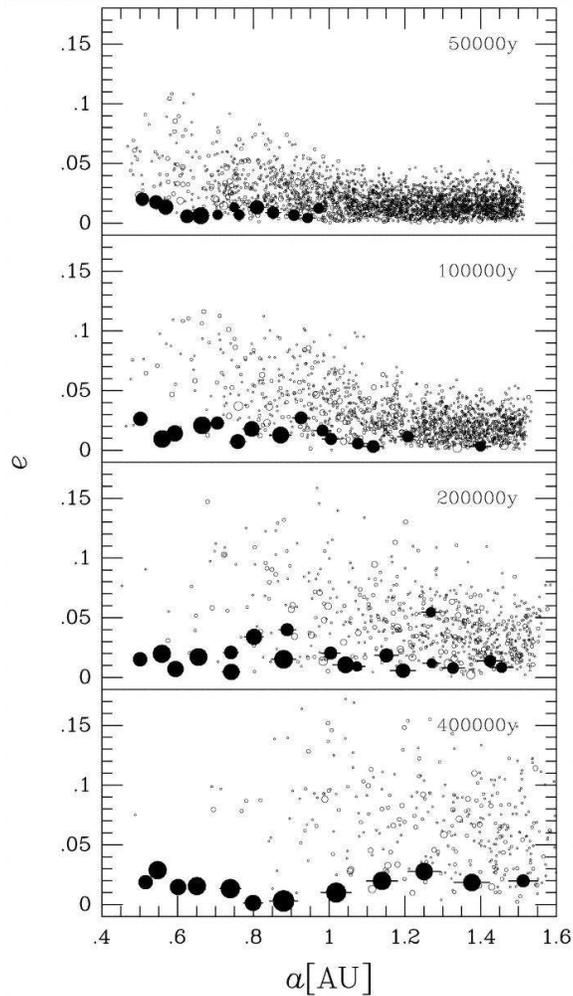} 
\caption{Snapshots in orbital eccentricity $e$ vs. semimajor axis $a$ in
simulations of the growth of planetary embryos \citep{Kokubo2002}.  The
radius of each particle is proportional to the simulation radius but is not to
scale on the x axis.This figure was cited from Kokubo and Ida,  Astrophys. J. 581, 666. (2002)  with copyright permission from IoP publishers.}
\label{fig:embform}
\end{center}
\end{figure}

Figure~\ref{fig:embform} shows snapshots in time of a simulation of
the formation of planetary embryos from planetesimals near 1 AU
\citep{Kokubo2002}. Accretion proceeds faster in the inner disk, such that the
outer disk is still dominated by planetesimals when embryos are fully-formed
in the inner disk. Oligarchic growth tends to form systems of embryos with
roughly comparable masses and separations of 5-10 mutual Hill radii
\citep{Kokubo1998, Kokubo2000, Weidenschilling1997}. The details of the embryo
distribution depend on the total mass and surface density distribution of the
disk \citep{Kokubo2002}. Typical embryo masses in a solar nebula model are a
few percent of an Earth mass, i.e., roughly lunar to Mars-sized
\citep{Kokubo2000, Collins2007}. Figure~\ref{fig:leinhardt_fig2} shows nine
distributions of embryos with a range in surface densities exponents $\alpha$
and surface densities $\Sigma_1$ (see Eqn. 1 of Leinhardt and Richardson\citep{Leinhardt2005}). For
surface density profiles steeper than $r^{-2}$, the embryo mass decreases with
orbital distance. Embryo masses scale roughly linearly with the local disk
mass, and formation times are much faster for more massive disks.

The process of embryo formation via runaway and oligarchic growth has very
recently come into question for three reasons.  First, disk turbulence
increases the random velocities of planetesimals, often above the critical
disruption threshold for km-sized planetesimals.  The
capacity of planetesimals to survive collisions is represented in terms of
of $Q_D^*$, the specific energy required to gravitationally disperse half of
the object's mass \citep{Melosh1997,Benz1999}.  For collisions more energetic
than $Q_D^*$, collisions are erosive rather than accretionary making it
difficult for embryos to grow.  In the presence of MRI(magneto-rotational instability) -driven turbulence \citep{Pessah2007},
accretionary growth of large bodies appears to require that larger bodies with
higher $Q_D^*$ already exist \citep{Ida2008}.  The critical size of these
large bodies is 300-1000 km.  Second, new collision models suggest that
planetesimals are weaker than previously estimated, such that accretion
requires either very slow collisions or pre-seeding of the disk with larger
objects \citep{Stewart2009}.  Third, statistical models that attempt to
reproduce the asteroid belt's observed size distribution must also resort to
seeding the region with large objects of at least 100 km in size
\citep{Morbidelli2008}.  These three lines of evidence all suggest that large,
100-1000 km bodies may have been required for the accretionary growth of the
much larger embryos.  This paradox could be resolved if planetesimals form via
the turbulent concentration plus gravitational collapse model of Johansen et al. (2007)\citet{Johansen2007}, who inevitably formed 1000 km-scale bodies in
MRI-turbulent disks.

\section{Late-Stage Growth of the Terrestrial Planets}

The planetary embryos formed during the previous oligarchic growth phase begin
to perturb one another once the local mass in planetesimals and embryos is
comparable \citep{Kenyon2006}.  The orbital eccentricities of embryos become
excited, which leads to a phase of close encounters and collisions with
moderate velocities.  Thus begins the final stage of terrestrial planet
formation, which ends with the formation of a few massive planets.
\citep{Wetherill1990,Wetherill1996,Chambers1998,Agnor1999,Chambers2001}. The
duration of this phase is shortened through the presence of Jupiter, which
increases the eccentricities of the embryos' orbits and hence the mutual
collision rates.

Wetherill (1992)\citet{Wetherill1992} was the first to suggest that the formation of planetary
embryos was not necessarily limited to the terrestrial planet region. He
proposed that planetary embryos formed also in the asteroid belt. The mutual
perturbations among the embryos, combined with the perturbations from Jupiter,
would have eventually removed all the embryos from the asteroid belt, leaving
in that region only a fraction of the planetesimal population on dynamically
excited orbits. For this reason, some of the simulations of Chambers and Wetherill  (1998)
\citet{Chambers1998} started with a population of embryos ranging from
$\sim$0.5 to $\sim$4 AU.

Most recent simulations take advantage of fast symplectic integrators such as 
{\tt Mercury}\citep{Chambers1999} or {\tt SyMBA}\citep{Duncan1998}. These 
integrators are optimized for planetary studies, and employ algorithms that allow for roughly  10 times fewer time steps per orbit as compared with a brute-force N-body 
integrator, for the same accuracy. These integrators also allow for close 
encounters between bodies, either by numerically solving the interaction 
component of the Hamiltonian ({\tt Mercury}) or by recursively subdividing the 
time step ({\tt SyMBA}). When performing integrations with these codes, it is 
always important to choose a time step that is small enough to resolve the 
orbits of the innermost particles with at least $\sim$ 20 time steps per orbit 
to avoid numerical errors \citep{Rauch1999, Levison2000}.  Collisions are
generally modeled in a very simplistic fashion, as inelastic  mergers occurring
anytime two bodies touch. Although this assumption appears  absurd, it has been
shown to have little to no effect on the outcome of  accretion simulations
\citep{Alexander1998}. However, more complex models  show that dynamical
friction from collisional debris may play an important  role at late stages
\citep{Levison2005}. 

A convenient approximation is often made to reduce the run time needed per
simulation, by neglecting graviational interactions between planetesimals
(see Raymond et al. (2006)for a discussion of this issue)\citep{Raymond2006}. Assuming
that planetesimals do not interact with each other, the run time $\tau$
scales with the number of embryos $N_e$ and the number of planetesimals,
$N_p$, roughly as $\tau \sim N_e^2 + 2 N_e N_p$. The non-interaction of
planetesimals eliminates an additional $N_p^2$ term. Note that $\tau$ refers
to the computing time needed for a given timestep. The total runtime is $\tau$
integrated over all timesteps for all surviving particles. Thus, a key element
in the actual runtime of a simulation is the mean particle
lifetime. Configurations with strong external perturbations (e.g., eccentric
giant planets) tend to run faster because the mean particle lifetime is
usually shorter than for configurations with weak external perturbations.


Tree codes, which subdivide a group of particles into cells using an opening angle criterion, have the advantage over serial codes in that the run time scales with particle number $N$ as  $N log N$ rather than $N^2$.  They can be run in parallel on several CPUs to further reduce the runtime.  Tree codes have been used to study planetary dynamics, but to date are only useful in the regime of large $N$ ($N \gtrsim 10^4$;\citep{Richardson00}).  The reason for this is that a large amount of computational "overhead" is required to build the tree, such that for small $N$ more computing time is needed for building the tree, and if run in parallel, for communication between processors.  The break-even point between serial codes and tree codes, for example, is at $N \sim 1000$\citep{Raymond05}.   An advantageous hybrid method for large $N$ accretion simulations is to integrate particles' orbits with a parallel tree code until $N$ drops to about 1000, then switch to serial code for the rest of the simulation\citep{Morishima08}.

A common problem with the current generation simulations is that the final terrestrial
planets are on orbits that are too eccentric and inclined with respect to the
real orbits.  The orbital excitation is commonly quantified by the normalized
angular momentum deficit\citep{Laskar1997}:

\begin{equation} 
AMD = \frac{\sum_{j} m_j \sqrt{a_j} \left(1 - cos 
(i_j) \sqrt{1-e_j^2}\right)} {\sum_j m_j \sqrt{a_j}}, 
\end{equation} 

\noindent where $a_j$, $e_j$, $i_j$, and $m_j$ refer to planet $j$'s semimajor
axis, eccentricity, inclination with respect to a fiducial plane, and mass. The
$AMD$ of the Solar System's terrestrial planets is 0.0018. For comparison, the Chambers and Wetherill (1998) Model C simulations, each consisting of at most ~50 bodies extending out to 4 AU and assuming the present orbits of Jupiter and Saturn, yield a median AMD of -0.033. The Chambers (2001) simulations 21-24, each consisting of about 150 bodies and also assuming the present Jupiter and Saturn, have a median AMD of -0.0050. Those simulations only extended out to 2 AU, and it is likely that their AMD would be even higher if they were extended out to 4 AU (e.g., the Chambers and Wetherill (1998) Model C Simulations, which extend out to 4 AU, have a median AMD about 50\% larger than in their Model B simulations, which only extend to 1.8 AU).

The missing physics responsible for this mismatch between simulations and
constraints is an open subject of scientific debate. It has been proposed
\citep{Kominami2002,Kominami2004} that a remnant fraction of the primordial
nebula would have damped the eccentricities and inclinations of the growing
planets. In this case, however, the simulations typically form systems of
planets that are too numerous and too small.  The  work has been extended by including the effects of
secular resonance sweeping as the solar nebula dissipates \citep{Nagasawa2006, Thommes2008}. This both forces mergers to reduce the
number of final terrestrial planets to be comparable to our Solar
System, and shortens the growth timescale so that there is sufficient
nebular gas at the finish to damp the eccentricities to match those of
the Solar System terrestrial planets.The MHD
turbulence of the nebula might also alleviate the problem, enhancing the
probability that the proto-planets collide with each other and thus leading to
systems with a smaller number of larger planets \citep{Ogihara2007}.  A problem with both
of these scenarios, however, is that since they occur on the timescale
comparable to the existence of the nebular gas (a few to $\sim$10 Myr). This
is not consistent with the significantly longer formation timescales
inferred from isotopic chronometry of the Earth-Moon system \citep{Touboul2007}, discussed more extensively at the end of this section.

Another possible way to reconcile the simulation results with the
constraints is the inclusion of dynamical friction. Dynamical friction occurs
if embryos and proto-planets evolve among a population of small planetesimals
with a total mass comparable to the total mass of the embryos. A bi-modal
distribution of embryos and planetesimals such as this is the likely result of
oligarchic growth \citep{Kokubo1996,Kokubo1998}. Dynamical friction produces the
equipartition of the ``excitation" energy (e.g., related to velocity dispersion, in analogy
to the temperature of a gas) between gravitationally interacting
bodies: the smaller ones obtain higher relative velocities, and the larger
ones lower. The relative velocity of embryos (hence their
eccentricities and inclinations) will therefore be kept low by dynamical
friction. The simulation of a large number of small planetesimals is, of
course, very CPU-intensive. Thus simulations typically neglect the effect of
the small bodies, or include only a limited number of them, which are,
therefore, artificially too massive.

\begin{figure}
\begin{center}
\includegraphics[width=5in]{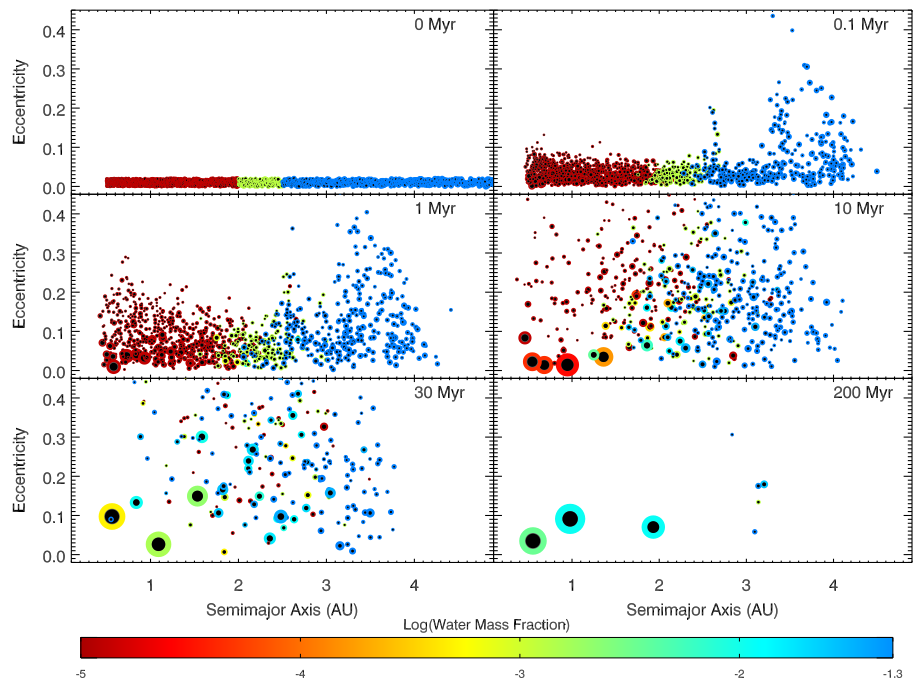} 
\caption{Snapshots in time from a simulation of the late-stage accretion of
terrestrial planets, starting from 1885 sub-isolation mass objects
\citep[]{Raymond2006}.  The size of each body is proportional to its
mass$^{1/3}$, the dark circle represents the relative size of each body's iron
core (in the black and white version, iron cores are shown only for bodies
larger than 0.05 \mearth), and the color corresponds to its water content (red
= dry, blue = 5\% water).  For a movie of this simulation, go to
http://casa.colorado.edu/$\sim$raymonsn and click on ``movies and graphics''.}
\label{fig:acc}
\end{center}
\end{figure}

\begin{figure}
\begin{center}
\includegraphics[width=5in]{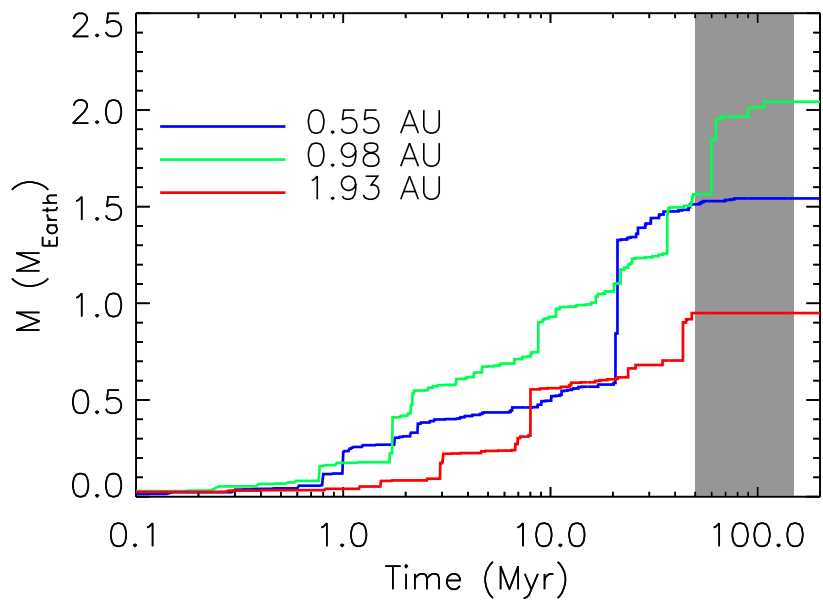} 
\caption{Growth of the three planets that formed in the simulation from
Fig.~\ref{fig:acc} \citep[]{Raymond2006}, labeled by their final orbital
distances.  The shaded region shows the constraint from isotopic measurements
for the timing of the Moon-forming impact \citep{Touboul2007}.}
\label{fig:mass-t}
\end{center}
\end{figure}

An example evolution of an accretion simulation is
shown in Figure~\ref{fig:acc} \citep{Raymond2006}.  This simulation started with 1886 sub-embryo
sized objects, and is one of the most computationally expensive to date, having
required 1.2 x 10{$^4$} CPU hours. The simulation contains a single Jupiter-mass
giant planet at 5.5 AU (not shown), and the evolution is characteristic of
simulations with low-eccentricity giant planets.  Eccentricities are excited
in the inner disk by mutual scattering between embryos, and in the outer disk
via resonant and secular forcing from the giant planet.  Dynamical friction
acts to keep the eccentricities of faster-growing embryos the smallest, and
accretion proceeds from the inside of the disk outward.  Only when embryos
reach a critical size can they scatter planetesimals and other embryos
strongly enough to cause large-scale radial mixing, which is evident in
Fig.~\ref{fig:acc} by the change in colors (which represent water contents) of
objects.  The Earth analog in this simulation started to accrete asteroidal
water only after $\sim$ 20 Myr of evolution, when it was more than half of its
final mass.  At the end of this simulation, three planets have formed: reasonable
Venus and Earth analogs at 0.55 and 0.98 AU, and a much-too-massive Mars
analog at 1.93 AU.  Figure~\ref{fig:mass-t} shows the growth of the three
planets in time.  The accretion of the Earth analog occurs on the correct
timescale, as it experiences its last giant impact at $t \approx 60$ Myr.  The
Venus and Mars analogs experience their final giant impacts at 22 and 40 Myr,
respectively.

The comparison between the results in Chambers and Wetherill (1998)\citet{Chambers1998} (no small bodies
included) and Chambers (2001)\citet{Chambers2001} (accounting for a bi-modal mass
distribution in the initial population) suggested that dynamical friction is
indeed important and can drive the simulation results in a good
direction. Thus O'Brien et al. (2006)\citet{OBrien06} performed new simulations, starting from a
system of 25 Mars-mass embryos from 0.5 to 4 AU, embedded in a disk of
planetesimals with the same total mass and radial extent as the
population of embryos, modeled with 1,000 particles. They performed two sets
of four simulations. In one set, called `EJS' for `Eccentric Jupiter
and Saturn,' Jupiter and Saturn are assumed to be initially on their current
orbits, and in the second set, called `CJS' for `Circular Jupiter and
Saturn,' they are assumed to be on nearly circular orbits with a smaller
mutual separation.

The results of the EJS simulations, with Jupiter and Saturn initially
on the current, eccentric orbits, can be compared directly to those of Chambers (2001)
\citet{Chambers2001}. The eccentricities and inclinations of the final
terrestrial planets \citep{Laskar1997, Chambers2001} measured through the AMD
turn out to be five times smaller, on average, than in Chambers' runs, and even
{$\sim$40\%} lower than in the real Solar System.  The median time for
the last giant impact is $\sim$30 Myr. For comparison, while Chambers (2001)
\citet{Chambers2001} does not report the time of last giant impact, his
Earth and Venus analogues take 54 and 62 Myr, respectively, to reach 90\% of
their final mass. (A recent study by one of the authors (SNR) and his colleagues suggest a spread of a factor of a few, and sometimes larger, between the last giant impact on Earth analogs in different simulations with the same set of initial conditions but different random number initializations).


The CJS simulations, with Jupiter and Saturn initially on
quasi-circular orbits and with smaller mutual separations, give a median time
for the last giant collision of about 100 Myr. They still give terrestrial planets that are a
bit too dynamically excited.  The eccentricities and inclinations, as
measured by the AMD, are about 60\% larger on
average than those of the real terrestrial planets. These somewhat
unsatisfactory results with regards to dynamical excitation do not
imply necessarily that Jupiter and Saturn had to have their current orbits when
the process of terrestrial planet formation started.  The spectacular
improvement in the results between the runs in Chambers (2001)\citet{Chambers2001} and the
EJS simulations in O'Brien et al. (2006)\citet{OBrien06} demonstrates the dramatic effect
of dynamical friction on reducing planetary excitation. With only 1,000
particles used to simulate the planetesimal disk, there is no reason to think
that the simulations by O'Brien et al. (2006)\citet{OBrien06} give a fully accurate
treatment of dynamical friction. Thus, it is possible that a future
generation of simulations, using more particles of smaller individual mass to
model the planetesimal disk, and allowing for the regeneration of
planetesimals when giant impacts occur between embryos\citep{Levison2005}, would treat dynamical friction more accurately and
lead to satisfactory results even with Jupiter and Saturn starting on circular
orbits.

Several statistical quantities exist to compare the properties of a
system of simulated terrestrial planets with the actual inner Solar System\citep{Chambers2001}.  These include the number and masses of the
planets, their formation timescales, the AMD of the system, and the radial concentration of the planets (the vast majority of the
terrestrial planets' mass is concentrated in an annulus between Venus and
Earth).  Reproducing all observed constraints in concert is a major goal of
this type of research\citep{Raymond2009}. 


With respect to formation timescales, constraints are available from measurement of radioactive isotopic systems in rocks on the Earth and Moon. To date these have yielded conflicting results. A very detailed analysis\citep{Touboul2007}, uses these chronometers, the identity of the tungsten isotopic ratios in the Moon and the Earth's mantle, and isotopic dating of the oldest moon rocks.  They conclude that the last giant impact--that which formed the Earth's Moon--occurred between 50-150 million years after the appearance of the first solids in the protoplanetary disk which formed the solar system. We assume--but cannot demonstrate-- that this giant impact did not occur a significant fraction of the Earth-formation time later than the collisions that built the Earth to its present size.  With that in mind, we argue that any simulations which grow the Earth on a timescale roughly between a few tens of millions and 150 million years are consistent with the indications from the geochemical data.

With respect to the radial distribution of terrestrial planet mass, the simulations described above start with a power law column density of solids. In contrast, Chambers and Cassen (2002)\citet{Chambers2002} simulated late-stage accretion by generating embryos from the detailed disk model of Cassen (2001)\citet{Cassen01} which contains a peak in the surface density at 2 AU (in that model, $\Sigma \propto a$ for $a <$ 2 AU, and $\Sigma \propto a^{-0.3}$ for $a >$ 2 AU).  They found that simulations with embryos generated from a standard MMSN model did a much better job of reproducing the properties of the terrestrial planets than the more detailed theoretical disk model.  Jin et al. (2008) \citet{Jin08} created a disk model with multiple zones, assuming that the ionization fraction of the gas varied radially, thereby affecting the local viscosity and causing pileups and dearths of gas at the boundaries between zones.  They suggested that non-uniform embryo formation in such a disk could explain Mars' small size.  Preliminary simulations by one of the authors (S.R.) and colleagues have called into question this suggestion. 

 
\subsection{Delivery of Water-Rich Material from the Asteroid Belt}

An outstanding application of the dynamical models is to the problem of the origin of Earth's water. The oceanic water content of the Earth is about 0.02\% the mass of the Earth, and various geochemical estimates put the total amount of water that was present early in the Earth's history at 5-50 times this number, some or all of which may yet reside in the mantle \citep{Abe00}. However, meteoritic evidence and theoretical modeling suggest that the protoplanetary disk at 1 AU was too warm at the time the gas was present to allow condensation of either water ice or bound water. Therefore, there has been a longstanding interest in models that deliver water ice or water-rich silicate bodies to the Earth during the latter's formation. Much of the isotopic and dynamical evidence against cometary bodies being a primary source, oft quoted in the literature, has been reviewed recently \citep{Lunine03}, and a comprehensive treatment of the geochemical evidence is beyond the scope of this review. Likewise, alternative models for local delivery of water, for example in the form of adsorbed water on nebular silicate grains \citep{Stimp08} have been proposed, but will not be described.  Of interest here is how the dynamical models described above can be used to quantify the  delivery of large bodies to the Earth from the asteroid belt, where chondritic material (in the form of meteorites) has an average $D/H$ ratio close to that of the Earth's oceans. 

The fact that the $ D/H$ ratio of Earth's water is chondritic prompted Morbidelli et al. (2000)\citet{Morby2000} to look at dynamical models of terrestrial planet formation
to investigate whether a sufficiently large amount of mass could be accreted
from the asteroid belt. They used simulations from Chambers and Wetherill (1998)\citet{Chambers1998}, in which planetary embryos beyond 2 AU had several times
the mass of Mars, and two new simulations with a larger number of individually
smaller embryos (masses ranging from a lunar mass at $\sim$1 AU to a Mars mass
at $\sim$4 AU).  They found that 18 out of the 24 planets formed in the
simulations accreted at least one embryo originally positioned beyond 2.5~AU.
When this happened, at least $\sim$10\% of the final planet mass was accreted
from this source. Assuming that the embryos originally beyond 2.5 AU had a
composition comparable to that of carbonaceous chondrites (namely with 5 to
10\% of mass in water), they concluded that these planets would be ``wet',
i.e.~they would start their geochemical evolution with a total budget of about
10 ocean masses of water or more. Moreover, Morbidelli et al. (2000)\citet{Morby2000} also
studied the evolution of planetesimals under the influence of the
embryos. They found that planetesimals from the outer asteroid belt also
contribute to the delivery of water to the forming terrestrial planets, but at
a considerably minor level with respect to the embryos. They also found that
comets from the outer planet region could bring no more than 10\% of an ocean
mass to the Earth, because the collision probability of bodies on cometary orbits with the earth is so low. From all these results, they concluded that the accretion
of a large amount of water is a stochastic process, depending on whether
collisions with embryos from the outer asteroid belt occur or not. Thus, they
envisioned the possibility that in the same planetary system some terrestrial
planets are wet, and others are water deficient.

The findings of Morbidelli et al. (2000)\citet{Morby2000} have been confirmed in a series of
subsequent works\citep{Raymond2004,Raymond2005,Raymond2006}. In particular, with simulations starting
from a larger number of smaller embryos, Raymond et al. (2007)\citet{Raymond2007} concluded that
the accretion of a large amount of water from the outer asteroid belt is a
generic result, and argued that the fact that 1/3 of the planets in Morbidelli et al. (2000)\citet{Morby2000} were dry was an artifact of small number statistics due
to the limited number of embryos used in those simulations.

\begin{figure}[]
\centering
\includegraphics[width=3in]{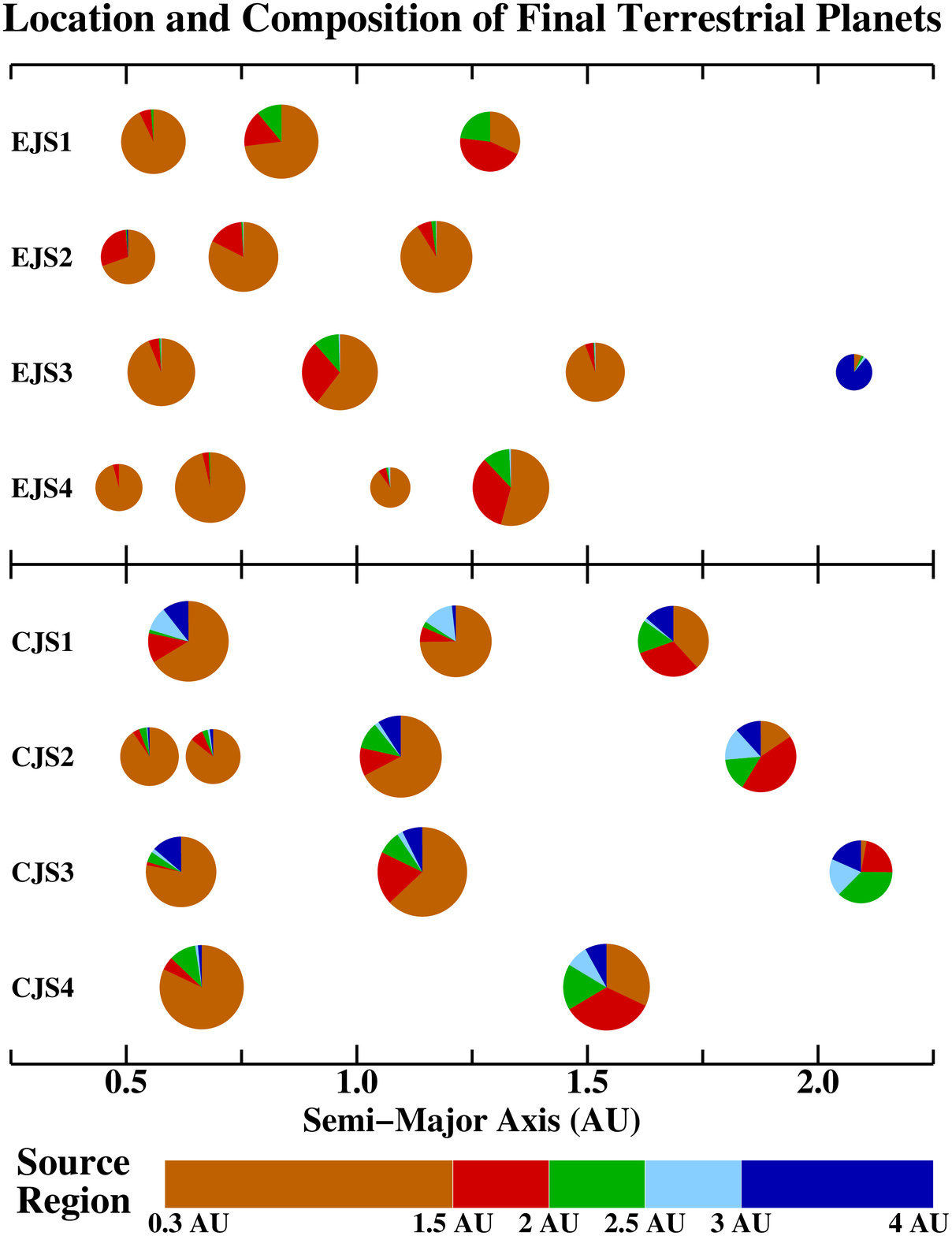}
\includegraphics[width=3in]{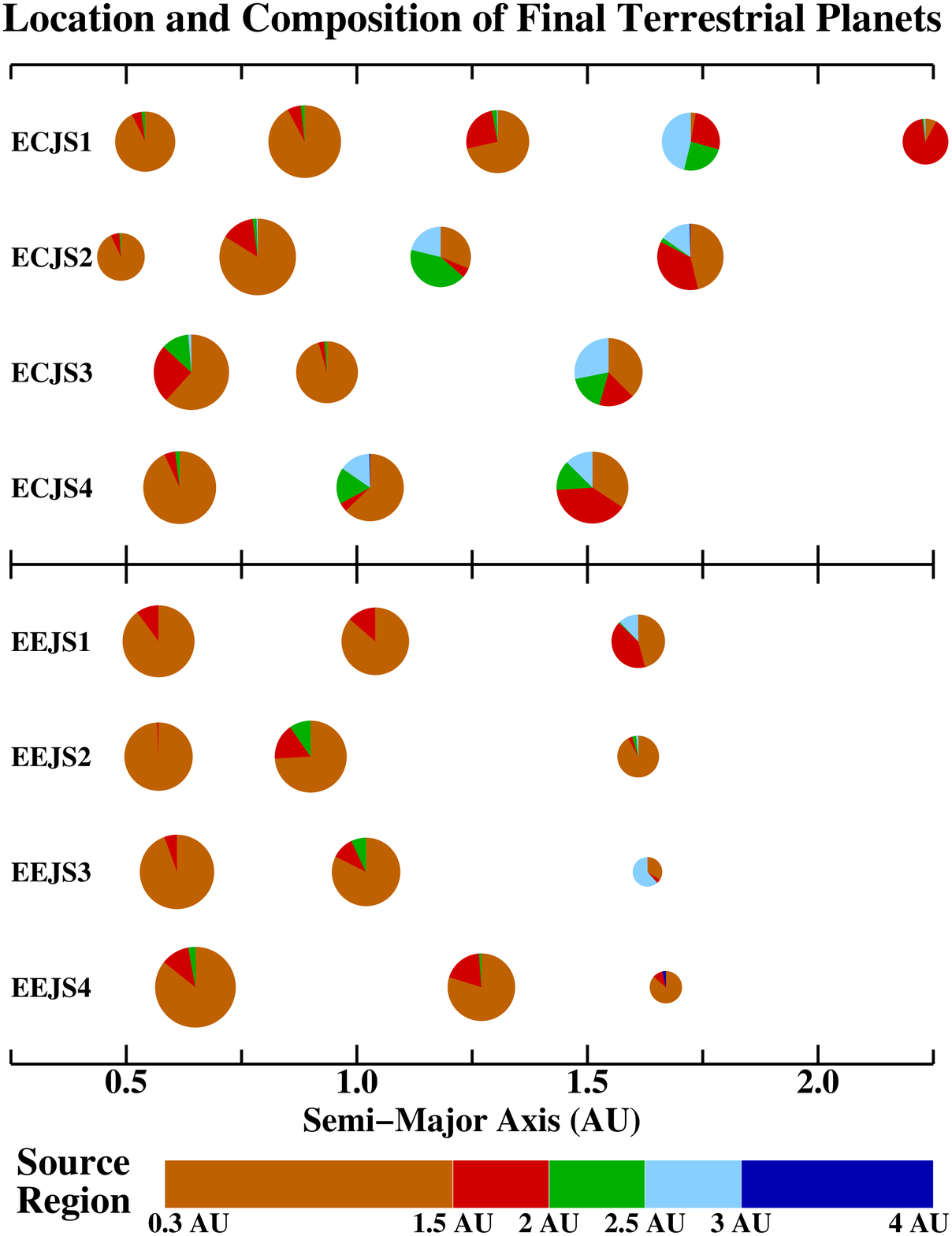}
\caption{Final terrestrial planets formed in the O'Brien et al. (2006) \citet{OBrien06} simulations
(EJS and CJS) as well as the ECJS and EEJS \citep{Raymond2009} simulations
discussed in the text.  Pie-diagrams show the relative contribution of material
from the different semi-major-axis regions, and the diameter of each symbol is
proportional to the diameter of the planet.\label{obrien}} 
\end{figure}

Figure ~\ref{obrien} shows the origin of the material incorporated in the final
terrestrial planets in the O'Brien et al. (2006)\citet{OBrien06} simulations. The top panel
concerns the set of four simulations with Jupiter and Saturn initially on
circular orbits, and the middle panel to the set with giant planets initially on
the current orbits. Each line refers to one simulation. Each planet is
represented by a pie diagram, with size proportional to the planet's diameter
and placed at its final semi-major axis.  The colors in each pie show the
contributions of material from the different semimajor-axis regions
shown on the scale at the bottom of the figure.  This represents the
feeding zone of each planet.  The feeding zones are not static, but generally widen
and move outward in time\citep{Raymond2006}.

An important difference is immediately apparent between the two sets of
simulations. In the set with Jupiter and Saturn initially on circular orbits,
an important fraction of the mass of all terrestrial planets comes from beyond
2.5 AU, and would likely be water-bearing carbonaceous
material. About 75\% of this mass is carried by embryos, the remaining part
by planetesimals. Thus, the idea that the water comes predominantly from the
asteroid belt is supported.  However, in the set of simulations with giant
planets initially on their current, eccentric orbits, none of the planets
accretes a significant amount material from beyond 2.5~AU. In that
case, if the asteroid belt is the source of water, it would have to
be through objects of ordinary chondritic nature, typical of its inner
part. We will come back to this idea below. This dramatic difference between
the cases with eccentric or circular giant planets had already been suggested\citep{Chambers2002}, and is explained by several authors\citep{Raymond2004},\citep{Raymond2006a}, and\citep{OBrien06}.

Thus, a crucial question for the origin of the Earth's water is
whether it is more reasonable to assume that the giant planets
initially had eccentric or circular orbits.  The core of a giant
planet is expected to form on a circular orbit because of strong damping by
dynamical friction and tidal interactions with the gas disk \citep{Kokubo1996,
Ward1993, Tanaka2004, Thommes2003}. 

Once an isolated giant planet is formed, if the mass is less than about
3~Jupiter masses, its interactions with the gas disk should not raise its
orbital eccentricit\citep{Kley2006} (but see Goldreich and Sari (2003)\citet{Goldreich2003}), but rather damp it out, if it is initially
non-zero. In our Solar System, however, we don't have an isolated giant
planet, but two. The dynamics of the Jupiter-Saturn pair has been
investigated\citep{Masset2001, Morbidelli2007, Pierens2008}. A typical evolution is that Saturn becomes locked into
the 2:3 resonance with Jupiter. This case is appealing because it may prevent
Jupiter from migrating rapidly towards the Sun, thus explaining why our Solar
System does not have a hot giant planet. Even in the case of 2:3
resonance locking, the orbital eccentricity of the giant planets
remain small. The eccentricity of Jupiter does not exceed 0.007.
There are, however, a few cases in which the
eccentricity of the giant planets can grow\citep{Morbidelli2007}. For instance, if a fast mass
accretion is allowed onto the planets, the resonance configuration can be
broken, and the eccentricity of Jupiter can temporarily grow to
$\sim$0.1. Also, if the planets are locked into the 3:5 resonance, the
eccentricity of Jupiter can be raised to 0.035, which is close to its current
value. All these cases, however, are unstable and temporary, so one has to
invoke the disappearance of the disk at the time of the excitation, otherwise
the planets would find another more stable configuration and the disk would
damp the eccentricities back to very small values. So, according to our
(limited) understanding of giant planet formation and gas-disk interactions, a
very small orbital eccentricity seems to be more plausible, but an eccentric
orbit cannot be ruled out with absolute confidence.

What seems more secure, conversely, is that when the terrestrial planet
formation process began, the orbits of Jupiter and Saturn had to have a
smaller mutual separation than their current value. In fact, all simulations
agree in showing that the interaction of the giant planets with the massive
planetesimal disk that would have existed in the early outer
Solar System leads to a significant amount of radial migration
\citep{Fernandez1984, Hahn1999, Gomes2004}. In particular, Saturn, Uranus and
Neptune migrate outwards, whereas Jupiter migrates inwards. Thus, the orbital
separation between Jupiter and Saturn grows with time.  Recently, a model of
the evolution and delayed migration of the giant planets has been
proposed, and it reproduces fairly well the current architecture of the outer
solar system\citep{Tsiganis2005,Gomes2005}. This ``Nice" model assumes that Jupiter
and Saturn were initially interior to their mutual 1:2 mean motion
resonance (MMR), and that the orbits of the giant planets at the time they
cross their 2:1 MMR were nearly circular.  
Giant planet migration and the crossing of the 2:1 resonance
could be delayed for hundreds of Myr, such that the initial configuration
would last for the entirety of the terrestrial planet formation process\citep{Gomes2005}.  The
second assumption of circular orbits at the time of the resonance crossing
does not dismiss, in principle, the simulations of terrestrial planet
formation starting with Jupiter and Saturn on eccentric orbits, because, in
these simulations, the giant planets eccentricities are damped very fast
by the ejection of material from the Solar System and meet the
requirements of Tsiganis et al. (2005)\citep{Tsiganis2005} and Gomes et al. (2005)\citep{Gomes2005} model
after a few tens of Myr. One should explain in this case, though, where
such eccentricity comes from. Conversely, the first assumption of a
smaller initial orbital orbital separation of Jupiter and Saturn is essential
for the success of that model.

For these reasons, we have performed a new set of four simulations, where
Jupiter and Saturn are assumed to have initially the current orbital
eccentricities, and an orbital separation consistent with the Tsiganis et al. (2005)\citep{Tsiganis2005}
and Gomes et al. (2005)\citep{Gomes2005} model.  The results in
terms of final eccentricities and inclinations of the terrestrial planets and
accretion timescales are intermediate between those of the two sets of
simulations in O'Brien et al. (2006)\citet{OBrien06} discussed above with AMD values
consistent with the Solar System values and a formation timescale consistent
with the Hf-W age of the Earth-Moon system\citep{Touboul2007}.  The origin
of the mass accreted by the terrestrial planets is presented in the bottom
strips of Fig.~\ref{obrien}.  The planets
at or beyond 1 AU, with only one exception, receive an important mass
contribution from the outer asteroid belt, that is comparable to, if not
larger, than that from the inner belt. The planets inside 1 AU typically do not
receive a significant mass contribution from the outer belt, and the
contribution from the inner belt is also very moderate.

We believe that we understand, at least at a qualitative level, the
differences between the results of these new runs and those of the set of O'Brien et al. (2006)
\citep{OBrien06} with Jupiter and Saturn on their current orbits,
in which none of the planets recieved significant outer-belt
material.  If the orbits of the giant planets are closer to each other, the
planets precess faster. Thus the positions of the secular resonances are
shifted outwards. In particular, the powerful $\nu_6$ resonance (occurring
when a body's perihelion precesses at the same rate as Saturn's), which is
currently at the inner border of the belt, moves beyond the outer belt. The
$\nu_6$ resonance can drive objects onto orbits with $e\sim 1$, such
that they are eliminated by collision with the Sun. It is therefore an
obstacle to the transport of embryos from the asteroid belt into the
terrestrial planet region. In fact, in decreasing their semi-major
axes from main belt-like values to terrestrial planets-like values, the
embryos in the EJS simulation have to pass through the resonance. Of
course, collisions with the growing terrestrial planets are also possible for
objects with a Main Belt-like semi-major axis and a large eccentricity, but
they are less likely.  An embryo can be extracted from the resonance by an
encounter with another embryo, but this is also an event with a moderate
probability. So, the flux of material from the belt to the terrestrial planet
region is enhanced if the $\nu_6$ resonance is not present.  This is the case
if the eccentricities of the planets are zero as in the CJS simulations (in
this case the resonance vanishes), or if the planets are closer to each other
as in the ECJS simulations (in which case the resonance is active, but it is
not between the terrestrial planets region and the asteroid belt).  In
the ECJS simulations, the $\nu_6$ is located around 3.4 AU. We stress that,
in order to move the $\nu_6$ resonance beyond the asteroid belt, it is not
necessary that Jupiter and Saturn are as close as postulated in the
 Tsiganis et al. (2005) \citep{Tsiganis2005} and Gomes et al. (2005) \citep{Gomes2005} model. The initial, less
extreme, orbital separations used in other models  \citep{Hahn1999, Gomes2004} 
would give a similar result.

We have recently performed several additional sets of simulations
\citep{Raymond2009}, including the EEJS (`Extra-Eccentric Jupiter and Saturn')
set. In four EEJS simulations, Jupiter and Saturn were placed at their current
semimajor axes but with starting eccentricities of 0.1. These systems
therefore experienced very strong perturbations from the $\nu_6$ resonance at
2.1 AU, which acted to remove material from the Mars region and also to
effectively divide the inner Solar System from the asteroid belt. These
simulations were the first to produce reasonable Mars analogs, but suffered in
terms of water delivery to the Earth. Scattering of embryos and planetesimals
during accretion decreased Jupiter and Saturn's eccentricities to close to
their current values, but the EEJS system does not allow for any delayed giant
planet migration as may be required by models of the resonant structure of the
Kuiper belt \citep{Malhotra2005, Levison2003b}. In fact, it is
important to note that the EJS simulations described above are absolutely
inconsistent with the Solar System's architecture because accretion damps the
eccentricities of Jupiter and Saturn to below their current values, and there
is no clear mechanism to increase them without affecting their semimajor
axes.

In conclusion, the simulations seem to support, from a dynamical
standpoint, the idea of the origin of water on Earth from the outer asteroid
belt. However, the stochasticity of the terrestrial planet accretion process,
the limitations of the simulations that we have used, and the uncertainties on
the initial configuration of the giant planets do not allow us to exclude a
priori the possibility that the Earth did not receive any
contribution from the outer asteroid belt, whereas it accreted an important
fraction of its mass from the inner belt or its vicinity.  For this reason, geochemical evidence has been used to try to constrain where the Earth's water came from. For example \citep{Right01} have argued that (a) oxygen isotopic differences and (b) siderophile element patterns limit the carbonaceous chondritic contribution to 1\% of the mass of the Earth.    Constraint (a) can be removed or relaxed if  the oxygen isotope composition of the Earth and the putative chondritic impactor were homogenized in the manner proposed for the Moon-forming impact event \citep{Pahlev09}.  (For the Moon-forming impactor such a process is deemed essential because the Earth and Moon have identical isotopic ratios for both oxygen and tungsten, whereas meteorites vary from these ratios).   Constraint (b) is a strong one only for relatively small bodies delivering water in a late veneer of material, or undifferentiated chondritic embryos mixing fully with the Earth's mantle during the main growth phase. If the embryo that delivered the water were differentiated then its core, containing most of the siderophile elements, would not mix with the Earth's mantle. 

\section{Extrapolation to Extrasolar Terrestrial Planet Systems}

What counts for terrestrial planet formation? The key parameters are 1) the
disk mass and radial density distribution, and 2) the giant planet properties
(mass, orbit, migration). Here we summarize some relevant issues
\citep[see][for a more detailed review]{Raymond2008}:

\begin{itemize} 

\item \textbf{Effect of Disk Properties} The accreted planet mass is slightly
more than linearly proportional to the disk mass because the planetary feeding
zone widens with disk mass due to stronger embryo-embryo scattering
\citep{Kokubo2006, Raymond2007b}. However, planets that grow to more than a
few Earth masses during the gaseous disk phase may accrete a thick H/He
envelope and be ``mini-Neptunes'' rather than ``super Earths''
\citep{Ikoma2000,Adams2008} . Whether such objects might be among the super-Earth mass planets observed around other stars is an interesting but as yet ill-constrained speculation. 

The disk's surface density profile is another key factor. For steeper density
profiles, the terrestrial planets form faster and closer to the star, are more
massive, more iron-rich and drier than planets that form in disks with
shallower density profiles \citep{Raymond2005}. Disks around other stars are
observed to have somewhat shallower density slopes\citep{Looney2003, Andrews2007a} than the $r^{-3/2}$
minimum-mass solar nebula model of Hayashi (1981)\citet{Hayashi1981} and Weidenschilling (1977)
\citet{Weidenschilling1977}. However, given
the preponderance of evidence that giant planets migrate, the validity of the
minimum-mass solar nebula for either our own solar system or other planetary systems is called into question \citep{Kuchner2004,
Desch2007}. Well-resolved observations of disk surface density profiles from facilities like ALMA will  help resolve this in the near future. 

\item \textbf{Low-Mass Stars}. Low-mass stars are in some sense an ideal place
to look for Earth-like planets, because an Earth-mass planet in the habitable zone
induces a stronger radial velocity signal in the star on a much shorter period
than for a Sun-like star \citep{Scalo2007, Tarter2007}.
However, sub-mm observations of the outer portions of dusty disks
around young stars show a roughly linear correlation between disk mass and
stellar mass, with a scatter of about 2 orders of magnitude in disk mass for a
given stellar mass \citep{Andrews2005, Andrews2007b,
Scholz2006}. Thus, low-mass stars tend to have low-mass disks which should
therefore form low-mass giant \citep{Laughlin2004} and terrestrial planets
\citep{Raymond2007b}. However, several low-mass stars are observed to host
massive (several Earth-mass), close-in planets \citep{Rivera2005,
Udry2007}.

\item \textbf{Effect of Giant Planet Properties}. Compared with a standard
case that includes giant planets exterior to the terrestrial planet forming
region, the following trends have been noted in dynamical simulations: 1) More
massive giant planets lead to fewer, more massive terrestrial planets
\citep{Levison2003, Raymond2004}; 2) More eccentric giant planets lead to
fewer, drier, more eccentric terrestrial planets \citep{Chambers2002,
Levison2003, Raymond2004, Raymond2006a, OBrien06}. Giant planets have a
negative effect on water delivery in virtually all cases, overly-perturbing
and ejecting much more water-rich asteroidal material than they allow to
slowly scatter inwards (S.  Raymond, unpublished data).

Hot Jupiter systems represent an interesting situation. In these systems, the
giant planet is thought to have formed exterior to the terrestrial planet zone,
then migrated through that zone \citep{Lin1996}. Recent simulations have shown
that the giant planet's migration actually induces the formation of rocky
planets in two ways: 1) interior to the giant planet, material is shepherded by
mean motion resonances, leading to the formation of very close-in terrestrial
planets \citep{Zhou2005, Fogg2005, Fogg2007, Raymond2006b, Raymond2008, 
Mandell2007}; and 2) exterior to the giant planet, the orbits of scattered
embryos are re-circularized by gaseous interactions leading to the formation of
a second generation of extremely water-rich terrestrial planets at $\sim$ 1 AU
\citep{Raymond2006b, Mandell2007}. Hence, a key factor is the chronology of
migration vs. disk dispersal. If the migration happens when there is still
a lot of mass in the disk for a good amount of time, then scattered
material can be saved and planets can formed.

\end{itemize}

\section{Conclusion}


Simulation of terrestrial planet formation has become a mature subfield of dynamical astronomy, with the potential to provide insight into the origin of our own solar system as well as that of the increasing number of multiple planet systems being discovered beyond our solar system. Further progress certainly will come from faster computers employing novelties such as, for example, many CPUs on a given chip allowing for easy communication between processors and improved performance and relevance of parallel codes. But additional insight into the physics and chemistry of the problem will be required as well. For example, while the general nature of our terrestrial
planet system seems to be broadly reproduced by the models, still unexplained is the presence of an embryo-sized body, Mars, in place of the more massive objects that the
simulations tend to yield. Are such outcomes common?  We cannot answer this question with the current state of maturity of the field. 


Another issue is the effect that collisions between embryos and the growing terrestrial planets have on the geochemistry of the latter. The challenge of quantifying in detail the chemical and physical processes that occur during giant impacts is a problem outside the scope of the dynamical modeling described here, but crucial in trying to relate the geochemistry of the Earth and other terrestrial planets to the source material from which they grew. Close collaboration between groups that specialize in these two very different types of numerical simulations may permit more detailed and confident geochemical predictions in the future.  And this, in turn, will increase our confidence in the predictions the models described herein can make for the properties of terrestrial planets around stars other than our own. 

\section{Acknowledgements}

We thank  Anders Johansen, Eiichiro Kokubo, and Zoe Leinhardt for their contributed figures. S.N.R. is grateful for funding from NASA's Origins of Solar Systems program (grant NNX09AB84G) and the Virtual Planetary Laboratory, a NASA Astrobiology Institute lead team, supported by NASA under Cooperative Agreement No. NNH05ZDA001C.  This paper is PSI Contribution 461.

\clearpage

\bibliographystyle{Science}
\bibliography{Lunine_et_al_dynamics_review}

\clearpage


\end{document}